\documentclass[a4paper,11pt]{article}

\usepackage[utf8]{inputenc}
\usepackage[english]{babel}
\usepackage{caption}
\usepackage{subcaption}
\usepackage{amsmath}
\usepackage{amsfonts}
\usepackage{amssymb}
\usepackage{amsthm}
\usepackage{graphicx}
\graphicspath{{Figures/}}
\usepackage{float}
\usepackage{verbatim}
\usepackage{psfrag}
\usepackage{geometry}
\usepackage{enumerate}
\usepackage{diagbox}
\usepackage{xcolor}
\usepackage{tabu}
\usepackage{cite}

\usepackage{fancyhdr}
\usepackage[auth-lg]{authblk}
\usepackage[square,numbers]{natbib}
\usepackage{hyperref}
\hypersetup{
    colorlinks=true,
    linkcolor=blue,
    filecolor=magenta,      
    urlcolor=cyan,
    pdfpagemode=FullScreen,
}
\usepackage{tikz}
\usepackage{relsize}
\numberwithin{equation}{section}

\newcommand{\be}{\begin{equation}}
\newcommand{\ee}{\end{equation}}
\newcommand{\bea}{\begin{eqnarray}}
\newcommand{\eea}{\end{eqnarray}}
\newcommand{\beano}{\begin{eqnarray*}}
\newcommand{\enano}{\end{eqnarray*}}

\title{\Large \bf Rotational Symmetry % Eigenstates 
	and Gauge Invariant Degeneracies \\
	on 2D Noncommutative Plane}
\author[1*]{\bf M.N.N.M. Rusli}
\author[2,3*]{\bf M.S. Nurisya}
\author[2,3]{\bf H. Zainuddin}
\author[4]{\bf M.F. Umar}
\author[5,6]{\bf A. Jellal}

\affil[1]{\it Department of Physics, Kulliyyah of Science (KOS), International Islamic University Malaysia (IIUM), 25200 Kuantan, Pahang, Malaysia}
\affil[2]{\it Laboratory of Computational Sciences and Mathematical Physics, Institute for Mathematical Research (INSPEM), Universiti Putra Malaysia, 43400 UPM Serdang, Selangor, Malaysia}
\affil[3]{\it Department of Physics, Faculty of Science, Universiti Putra Malaysia, 43400 UPM Serdang, Selangor, Malaysia}
\affil[4]{\it Faculty of Science and Mathematics, Universiti Pendidikan Sultan Idris, 35900 Tanjung Malim, Perak, Malaysia}
\affil[5]{\it Laboratory of Theoretical Physics,  
	Faculty of Sciences, Choua\"ib Doukkali University, PO Box 20, 24000 El Jadida, Morocco}
\affil[6]{\it Canadian Quantum  Research Center,
	204-3002 32 Ave Vernon, BC V1T 2L7,  Canada}
\date{\vspace{-1\baselineskip}{\textit{$^*$\small{corresponding email: nazrennazmi98@gmail.com, risya@upm.edu.my}}}}

\fancypagestyle{plain}{%
\fancyhf{} % clear all header and footer fields
\fancyhead[C]{\small{}}
\fancyfoot[C]{\small{\thepage}}

}

\fancypagestyle{fancy2}{
\fancyhf{} % clear all header and footer fields
\lhead{\small}

}

\begin{document}
\renewcommand{\abstractname}{\vspace{-3\baselineskip}}
\pagestyle{plain}
    \begin{@twocolumnfalse}
    \maketitle
	\begin{abstract} \noindent
	    \begin{comment}
	    \\ The solution to the eigenvalue problem involving isotropic harmonic oscillator in noncommutative phase space using the unitary irreducible representation of the non-central elements of the Lie algrebra of the triply extended group of translations of $\mathbb{R}^4$ on $L^2\left(\mathbb{R}^2, dr_1 dr_2\right)$ in symmetric gauge is solved. It has been observed that the system is generally restricted such that the product of the scaled magnetic field and noncommutativity has to obey the following condition; $0 \leq \mathbf{B}\theta \leq \hbar^2$ where $\mathbf{B} = \frac{B}{\hbar}$. Hence for $\mathbf{B}\theta = 0$ which directly implies that $\mathbf{B} = 0$, the energy spectrum of the system is equivalent to that of the system when it lies in the noncommutative space as opposed to noncommutative phase space where the coordinates of momenta commute with each other. As for $\mathbf{B}\theta = \hbar^2$, the system exhibits infinite degeneracy in its spectrum regardless of the values of the parameters used while in between both of extreme ends of the condition, the system's degeneracy is very unique to certain types of particle. If the condition of degeneracy is not satisfied, the system is nondegenerate while if it does, the system is atleast doubly degenerate. The similarity of the two cases above is the fact that $\mathbf{B}$ predominantly affects the system while $\theta$ barely produce any significant effect on the system.  
	    \end{comment}
		\\ We obtain the gauge invariant energy eigenvalues and degeneracies together with rotationally symmetric wavefunctions of a particle moving on 2D noncommutative plane subjected to homogeneous magnetic field $B$ and harmonic potential. This has been done  by using the phase space coordinates transformation based on 2-parameter family of unitarily equivalent irreducible representations of the nilpotent Lie group $G_{NC}$. We find that the energy levels and  states of the system are unique and hence, same goes to the degeneracies as well since they are heavily reliant on the applied $B$ and the noncommutativity $\theta $ of coordinates. Without $B$, we essentially have a noncommutative planar harmonic oscillator under generalized Bopp shift or Seiberg-Witten map. The degenerate energy levels can always be found if $\theta$ is proportional to the ratio between $\hbar$ and $m\omega$. For the scale $B\theta = \hbar$, the spectrum of energy is isomorphic to Landau problem in symmetric gauge and hence, each energy level is infinitely degenerate regardless of any values of $\theta$. Finally, if $0 < B\theta < \hbar$, $\theta$ has to also be proportional to the ratio between $\hbar$ and $m\omega$ for the degeneracy to occur. These proportionality parameters are evaluated and if they are not satisfied then we will have non-degenerate energy levels. Finally, the probability densities and effects of $B$ and $\theta$  on the system are properly shown for all cases.
		\vspace{1cm}
		
	\end{abstract}

	\end{@twocolumnfalse}

\thispagestyle{fancy2}

\newpage
\section{Introduction}\label{introduction}
Historically, the notion of noncommutative structure in spacetime coordinates primarily predates back to Heisenberg and formally realized by Snyder in 1947 \cite{history1} as an effective way to avoid or at least, ameliorate the short-distance singularities which had plagued quantum field theory and in particular, gauge theories in the early days. However, this idea only became a subject of interest for a short while due the remarkable success of the ensuing emergence of renormalization scheme. We refer to \cite{review1,review2,review3} to gain some insight on the historical context and pertinent reviews in greater details. Nonetheless in recent years, the research involving these noncommuting coordinates has gained back its momentum due to the discovery of its applicability in the framework of superstring theories and of quantum gravity \cite{review4}. For some reviews of these topics and in particular, of quantum field theory on noncommutative spacetime, we refer to \cite{review1,review5}.

\begin{comment}
Nonetheless, this idea did not gain much attention in the beginning due to the success of the emergence of renormalization formalism at the time which proved to be successful at accurately predicting numerical values for physical observables. In the 1980s, this

Recently, noncommuting coordinates have regained interest, in particular in the framework of superstring theories and of quantum gravity. Roughly speaking, the intrinsic length scale of the strings induces a noncommutative structure on space-time at low scales \cite{review4}. For further readings on topics related to quantum field theory on noncommutative space-time and its renormalization, see the following references \cite{review5,source13,source14,source15,source16}. 

In this paper we restrict ourselves to the version of non-commutative quantum mechanics which describes a quantum system with two degrees of freedom with its own non-canonical commutation relations as will be pondered later on.
\end{comment}

There are numerous works which tackle the issue of the one particle sector of noncommutative field theories i.e, noncommutative quantum mechanics (NCQM) in different settings. This includes harmonic oscillator \cite{harmonic1,harmonic2,source3}, magnetic field \cite{Dayi1,Dayi2,Jellal2002,Jellal2003,magnetic}, hydrogen atom \cite{hydrogen1}, central potential \cite{central,my1stpaper}, Landau problem \cite{Landau}, Klein-Gordon and Dirac oscillators \cite{KleinGordonDirac1,KleinGordonDirac3,Dirac1}, Aharonov-Casher effect\cite{Mirza2} and the list goes on. 
One of the authors \cite{Jellal2021}  studied
a system of spinless electrons moving in a 2D
noncommutative space in the presence of a perpendicular magnetic field  and confining
harmonic potential. His focus was on  the orbital magnetism of the
electrons in different regimes of temperature, magnetic field and
noncommutative parameter $\theta$. In fact, he proved that the degeneracy of Landau levels
can be lifted by the $\theta$-term appearing in the electron energy spectrum at weak
magnetic field. 

%Using the Berezin–Lieb inequalities for thermodynamical
%potential, it is shown that in the high-temperature limit, the system exibits
%a magnetic θ -dependent behaviour, which is missing in the commutative case.
%Moreover, a correction to susceptibility at low T is observed. Using the Fermi–
%Dirac trace formulas, a generalization of the thermodynamical potential, the
%average number of electrons and the magnetization is obtained. There iscritical point where the thermodynamical potential becomes infinite in both of
%the two methods above. So at this point we deal with the partition function by
%adopting another approach. The standard results in the commutative case for
%this model can be recovered by switching off the θ -parameter.

In the above mentioned works, the center of interest usually revolves around eigenvalue problem which most of the time are treated using algebraic method while the analytical approach receives relatively lesser attention in comparison which hinders us from gaining some direct physical insight. Apart from that, the sources which discuss the degeneracies of NCQM models are also noticeably scarce and face the issue of gauge dependency (e.g, in \cite{source1,DulatHall}). Therefore, these concerns shall be our main focus to be addressed in this paper particularly for charged quantum harmonic oscillator on noncommutative plane. We will not however be utilizing the minimal coupling prescription as is done naively on many occasions in the literature as it yields gauge dependence of the underlying energy spectra e.g, in anisotropic harmonic oscillator and quantum Hall effect \cite{Dayi1,Jellal2003,source2}. We will instead rely on families of self-adjoint irreducible representations of the universal enveloping algebra $\mathcal{U}(\mathfrak{g}_{NC})$ of the Lie algebra $\mathfrak{g}_{NC}$ as the kinematical symmetry group of 2D noncommutative quantum mechanics whose Lie algebra has been established in \cite{group} that is able to produce gauge invariant spectra.

The paper is organized as follows: In Section 2, we briefly revisit the minimal coupling prescription and state the alternative gauge invariant phase space coordinates transformation to be used later. Then, the time-independent Schr\"{o}dinger equation of an isotropic harmonic oscillator in noncommutative phase space is solved to express the energy eigenvalues and eigenstates that emerge. We rely on the symmetric part of the 2-parameter (r, s) family of irreducible self-adjoint representations $\mathcal{U}(\mathfrak{g}_{NC})$ as the coordinate transformation. It turns out that the product of the two parameters of the system i.e, magnetic field and noncommutativity has to be $0 \leq B\theta \leq \hbar$ so that the effective mass that characterizes the system is real which then gives rise to three possible cases. As a result, the third section is dedicated to the study of the first case i.e, in the absence of magnetic field ($B\theta = 0$). The gauge invariant energy eigenvalues and degeneracies are evaluated and the rotationally symmetric eigenstates are defined. We also provide an instructive example. The similar discussion are done in the forth section as in the previous one for the remaining two cases in the presence of homogeneous magnetic field for $0 < B\theta < \hbar$ and $B\theta = \hbar$ respectively. The final section is allocated to serve as visual guide on the probability densities and effect of magnetic field and noncommutativity on the system for all cases.

\section{Energy spectrum on noncommutative plane}\label{sec:2}
Before we begin our discussion on the gauge invariant coordinates transformation, we will briefly revisit the so called minimal coupling prescription which is used in many literature. In standard quantum mechanics, the quantum phase space coordinates comprise of the Hermitian operators $\hat{x}$,$\hat{y}$,$\hat{p}_x$ and $\hat{p}_y$ defined on $L^2(\mathbb{R}^2,dxdy)$ with the following commutation relations
\begin{align}
&\left[\hat{x},\hat{y}\right] =  \left[\hat{p}_x,\hat{p}_y\right] = 0,\\
& 
\left[\hat{x},\hat{p}_x\right] = \left[\hat{y},\hat{p}_y\right] = i\hbar\mathbb{I},\\
& \left[\hat{x},\hat{p}_y\right] = \left[\hat{y},\hat{p}_x\right] = 0,\label{eq:2.7}
\end{align}
where $\mathbb{I}$ is the identity operator on $L^2 \left(\mathbb{R}^2, dr_1 dr_2\right)$. The above commutation relations correspond to the 5-dimensional Weyl-Heisenberg group, $G_{WH}$. We can try to incorporate noncommuting spatial coordinates governed by $[\hat{X},\hat{Y}] = i\theta\mathbb{I}$ where $\theta$ is a real frame dependent parameter. The minimal coupling prescription in noncommutative quantum mechanical problem in the presence of a constant magnetic field is derived from the following gauge potential
\begin{align}
    \mathbf{\hat{A}} (\hat{X},\hat{Y}) = (-B(1-\alpha)\hat{Y},\alpha B\hat{X}), 
\end{align}
where Landau and symmetric gauges correspond to $\alpha= 1$ and $\mathlarger{\alpha = \frac{1}{2}}$ respectively. Later, this prescription can be used naively to write down the kinematical momentum operators as follows
\begin{align}
    \hat{P}_i = \hat{p}_i - e\hat{A}_i,\qquad i = x,y.
\end{align}
Subsequently, the above transformation can be used to map the noncommutative spatial coordinates to commutative ones by the well known generalized Bopp shift or Seiberg-Witten map
\begin{align}
 &   \hat{X} = \hat{x} - \frac{\theta}{2\hbar}\hat{p}_y,\\
 & \hat{Y} = \hat{y} + \frac{\theta}{2\hbar}\hat{p}_x,\\
 &
    \hat{P}_x = \hat{p}_x,\\
    & \hat{P}_y = \hat{p}_y.
\end{align}
In \cite{source2}, this noncommutative setup has been shown explicitly to yield gauge dependency via eigenfrequencies of the underlying energy spectra for the cases of anisotropic harmonic oscillator and quantum Hall effect which is inconsistent in the context of noncommutative quantum mechanics. Therefore in this work, we will be counting on families
of self-adjoint irreducible representations of the universal enveloping algebra $\mathcal{U}(\mathfrak{g}_{NC})$ of the Lie algebra $\mathfrak{g}_{NC}$ whose corresponding Lie group $G_{NC}$ has been established in an earlier paper \cite{group} which is the kinematical symmetry group of 2D NCQM. We will not delve any further into the group theoretical structure as that has been done in \cite{source2}. What we are interested in is using the result of gauge invariant coordinates transformation in the paper i.e, 2-parameter family of equivalent self-adjoint irreducible representation of the universal enveloping algebra $\mathcal{U}(\mathfrak{g}_{NC})$ on the smooth vectors of $L^2(\mathbb{R}^2, dxdy)$, a family to which Landau and symmetric gauge representations belong where it is formulated as below
\begin{align}
 &   \hat{X}^s = \hat{x} - s\frac{\theta}{\hbar}\hat{p}_y,\\
 &
    \hat{Y}^s = \hat{y} + (1-s)\frac{\theta}{\hbar}\hat{p}_x,
\\
&
    \hat{\Pi}_x^{\alpha,s} = \frac{(1 - \alpha)\hbar B}{\hbar - \alpha\theta B}\left(\hat{y} - \frac{s\theta}{\hbar}\hat{p}_x \right) + \hat{p}_x, \\
  &  \hat{\Pi}_y^{\alpha,s} = -\alpha B\left[\hat{x} + \frac{(1-s)\theta}{\hbar}\hat{p}_y\right] + \hat{p}_y, 
\end{align}
where %$ (\alpha=1,s=0) $ correspond to 
the Landau and 
 %$\left(\alpha = \frac{\hbar}{\hbar + \sqrt{\hbar^2 - \hbar B \theta}}, s = \frac{1}{2}\right)$ for
symmetric gauges correspondent to, respectively,
\begin{align}
&   r = 1,\quad s = 0, %,\qquad \mbox{for Landau gauge,}\nonumber
\\
&
    r = \frac{\hbar}{\hbar + \sqrt{\hbar^2 - \hbar B \theta}},\qquad s = \frac{1}{2}. %,\qquad \mbox{for symmetric gauge}.
\end{align}
In this work, we will be focusing on the symmetric gauge part since the energy eigenvalues and the associated degeneracies that will be obtained can naturally be extended to Landau gauge as well due to gauge invariance. For the wavefunctions, the mathematical structure should only apply to the symmetric gauge and also any other gauge choices with underlying rotational symmetry under simple substitution.

The Hamiltonian of a particle of mass $m$ which oscillates with an angular frequency $\omega$ under the influence of an isotropic harmonic oscillator potential in the noncommutative phase space can be denoted as
\begin{align} \label{eq:2.1}
\hat{H} = \frac{1}{2m} \left(\hat{\Pi}_1^2 + \hat{\Pi}_2^2\right) + \frac{1}{2} m\omega^2 \left(\hat{Q}_1^2 + \hat{Q}_2^2\right).
\end{align}
By representation theory, the corresponding gauge invariant (symmetric gauge part) phase space coordinates transformation between noncommutative operators and commutative ones based on 2-parameter family of equivalent self-adjoint irreducible representation of the universal enveloping algebra $\mathcal{U}(\mathfrak{g}_{NC})$ on the smooth vectors of $L_2(\mathbb{R}^2,dxdy)$ can be obtained as follows \cite{source2}
%\begin{equation}
%\hat{X} = \hat{x} - \frac{\vartheta}{2\hbar}\hat{p}_y, \nonumber
%\end{equation}
\begin{align}
&\hat{X} = \hat{x} - \frac{\theta}{2\hbar}\hat{p}_y, \label{eq:2.2a}\\
& \hat{Y} = \hat{y} + \frac{\theta}{2\hbar}\hat{p}_x, 
\\
&
\hat{\Pi}_{x} =  \frac{\hbar B}{\hbar + \sqrt{\hbar(\hbar - B\theta)}} \hat{y} + \frac{\hbar + \sqrt{\hbar(\hbar - B\theta)}}{2\hbar} \hat{p}_x, \\
&\label{eq:2.2}
\hat{\Pi}_{y} = - \frac{\hbar B}{\hbar + \sqrt{\hbar(\hbar - B\theta)}} \hat{x} + \frac{\hbar + \sqrt{\hbar(\hbar - B\theta)}}{2\hbar} \hat{p}_y,
\end{align}
where $\theta$ is a small, positive parameter which measures the additionally introduced noncommutativity between the observables of the two spatial coordinates. The self-adjoint differential operators on the space of smooth vectors of $L^2(\mathbb{R}^2)$ obeys the following commutation relations
%\begin{equation}\nonumber
%\left[\hat{X},\hat{Y}\right] = i\vartheta\mathbb{I},
%\end{equation} 
\begin{align}
&\left[\hat{X},\hat{Y}\right] = i\theta\mathbb{I}, \\
& \left[\hat{\Pi}_x,\hat{\Pi}_y\right] = i\hbar B\mathbb{I},
\\
&
\left[\hat{X},\hat{\Pi}_x\right] = \left[\hat{Y},\hat{\Pi}_y\right] = i\hbar\mathbb{I}, \\
& \left[\hat{X},\hat{\Pi}_y\right] = \left[\hat{Y},\hat{\Pi}_x\right] = 0,
\end{align}
%\begin{equation}
%\left[\hat{X},\hat{\Pi}_y\right] = \left[\hat{Y},\hat{\Pi}_x\right] = 0,
%\end{equation}
where $\mathbb{I}$ being the identity operator on $L^2\left(\mathbb{R}^2, dr_1 dr_2\right)$. Note that the magnetic field i.e, $ B $ can be rescaled
$B \rightarrow \mathlarger{\frac{eB}{c}}$ to connect our notation with the usual literature on Landau problem. In \cite{source2}, the cyclotron frequency was defined to be 
%\begin{align} \label{cyc_freq}
    $\omega_c = \frac{B}{m}.$
%\end{align}
So strictly speaking, in SI units, due to the absence of the charge of particle of interest in %\eqref{cyc_freq}
$\omega$, we can consider that it is already being absorbed in $B$. Therefore, throughout this work, $B$ will be defined as the magnetic field multiplied by a unit charge. However, we will simply call it magnetic field in the sequel.
\begin{comment}
The momentum operators can be rewritten as follows
\begin{align} \label{eq:2.4}
\hat{p}_x & = -i\hbar\frac{\partial}{\partial x}, \hat{p}_y = -i\hbar\frac{\partial}{\partial y}, \nonumber
\\
\hat{p}_x^2 & = -\hbar^2\frac{\partial^2}{\partial x^2}, \hat{p}_y^2 = -\hbar^2\frac{\partial^2}{\partial y^2},
\end{align}
and
\begin{equation}
\hat{L_z} \equiv -i\hbar\left(x\frac{\partial}{\partial y} - y\frac{\partial}{\partial x}\right),
\end{equation}
is the z-component of the orbital angular momentum operator. 
\end{comment}

By substituting from \eqref{eq:2.2a} until \eqref{eq:2.2} into \eqref{eq:2.1}, the Hamiltonian is of the form
\begin{align}
	\begin{aligned}
\hat{H} = & \frac{1}{2m} \left(\frac{\hbar B}{\hbar + \sqrt{\hbar(\hbar - B\vartheta)}} \hat{y} + \frac{\hbar + \sqrt{\hbar(\hbar - B\vartheta)}}{2\hbar} \hat{p}_x\right)^2 
\\
& + \frac{1}{2m} \left(- \frac{\hbar B}{\hbar + \sqrt{\hbar(\hbar - B\vartheta)}} \hat{x} + \frac{\hbar + \sqrt{\hbar(\hbar - B\vartheta)}}{2\hbar} \hat{p}_y\right)^2 
\\
& + \frac{1}{2} m\omega^2 \left(\hat{x} - \frac{\vartheta}{2\hbar}\hat{p}_y\right)^2 + \frac{1}{2} m\omega^2 \left(\hat{y} + \frac{\vartheta}{2\hbar}\hat{p}_x\right)^2.
	\end{aligned}
\end{align}
After a few algebraic manipulation steps, then
\begin{align} \label{eq:2.9}
		\begin{aligned}
\hat{H} = &  \left(\frac{2\hbar^2 + 2\hbar\sqrt{\hbar^2 - \hbar B\theta} - \hbar B\theta + m^2\omega^2\theta^2}{8m\hbar^2}\right)\left(\hat p_x^2 + \hat p_y^2\right) 
\\
& + \left(\frac{\hbar^2 B^2 + m^2\omega^2(2\hbar^2 + 2\hbar\sqrt{\hbar^2 - \hbar B\theta} - \hbar B\theta)}{2m(2\hbar^2 + 2\hbar\sqrt{\hbar^2 - \hbar B\theta} - \hbar B\theta)}\right)\left(\hat{x}^2 + \hat{y}^2\right)
 - \left(\frac{\hbar B + m^2\omega^2\theta}{2\hbar m}\right)\hat{L}_z.
	\end{aligned}
\end{align}
where $L_z$ is the z-component of the orbital angular momentum. By rearranging \eqref{eq:2.9}, we can introduce a new effective mass and frequency so that the expression above excluding the $\hat{L}_z$-term can have the form of the Hamiltonian of a planar isotropic harmonic oscillator as follows
\begin{equation} \label{eq:2.10}
\hat{H} = \frac{1}{2M} \left(\hat{p}_x^2 + \hat{p}_y^2\right) + \frac{1}{2} M\Omega^2 \left(\hat{x}^2 + \hat{y}^2\right) - \gamma \hat{L_z}.
\end{equation}
Hence, by comparing \eqref{eq:2.9} and \eqref{eq:2.10}, the effective mass is
\begin{align} \label{eq:2.11}
    M & = \frac{4m\hbar^2}{2\hbar^2 + 2\hbar\sqrt{\hbar^2 - \hbar B\theta} - \hbar B\theta + m^2\omega^2\theta^2},
\end{align}
\begin{comment}
    \frac{1}{2M} & = \frac{2\hbar^2 + 2\hbar\sqrt{\hbar^2 - \hbar B\theta} - \hbar B\theta + m^2\omega^2\theta^2}{8m\hbar^2}, \nonumber
\end{comment}
whereas the effective frequency is
\begin{align}
    \Omega = \sqrt{\frac{2\hbar^2 + 2\hbar\sqrt{\hbar^2 - \hbar B\theta} - \hbar B\theta + m^2\omega^2\theta^2}{4m\hbar^2}},
    \label{omega}
\end{align}
and we have set  $ \gamma $ as
\begin{align}
\gamma= \frac{\hbar B + m^2\omega^2\theta}{2\hbar m}.
\label{gamma}
\end{align}		
\begin{comment}
    \frac{1}{2}M\Omega^2 & = \frac{\hbar^2 B^2 + m^2\omega^2(2\hbar^2 + 2\hbar\sqrt{\hbar^2 - \hbar B\theta} - \hbar B\theta)}{2m(2\hbar^2 + 2\hbar\sqrt{\hbar^2 - \hbar B\theta} - \hbar B\theta)},
    \\
    \Omega & = \sqrt{\frac{\hbar^2 B^2 + m^2\omega^2(2\hbar^2 + 2\hbar\sqrt{\hbar^2 - \hbar B\theta} - \hbar B\theta)}{m(2\hbar^2 + 2\hbar\sqrt{\hbar^2 - \hbar B\theta} - \hbar B\theta)}} \nonumber
\end{comment}
The corresponding stationary Schr\"{o}dinger equation is therefore
\begin{multline} \label{eq:2.14}
-\frac{\hbar^2}{2M} \left(\frac{\partial^2 \Psi(x,y)}{\partial x^2} + \frac{\partial^2 \Psi(x,y)}{\partial y^2}\right) + \frac{1}{2} M\Omega^2 (x^2+y^2)\Psi(x,y) - \gamma L_z\Psi(x,y) = E\Psi(x,y).
\end{multline}
Since the Hamiltonian \eqref{eq:2.10} is rotationally symmetric, it is appropriate to work in polar coordinates $ (r,\varphi) $. Then \eqref{eq:2.14} becomes
\begin{multline}
-\frac{\hbar^2}{2M} \left(\frac{\partial^2 \Psi(r,\varphi)}{\partial r^2} + \frac{1}{r}\frac{\partial \Psi(r,\varphi)}{\partial r} + \frac{1}{r^2}\frac{\partial^2 \Psi(r,\varphi)}{\partial \varphi^2}\right) + \frac{1}{2} M\Omega^2 r^2 \Psi(r,\varphi) - \gamma L_z\Psi(r,\varphi) = E\Psi(r,\varphi),
\end{multline}
where $L_z = -i\hbar\frac{\partial}{\partial\varphi}$.
After solving the above eigenvalue problem,
the expression of the eigenvalues are obtained and in terms of $\Omega$ and $\gamma$
\begin{align} \label{eq:2.16}
E_{n_r,m_l} & = \left(2n_r + |m_l| + 1\right)\hbar\Omega - m_l\hbar\gamma,
\end{align}
\begin{comment}
\begin{equation} 
\begin{split}
E_{n_r,m_l} & = \left(2n_r + |m_l| + 1\right)\hbar\Omega - m_l\hbar\gamma,
\\
& = \left(2n_r + |m_l| + 1\right)\hbar\sqrt{\frac{\hbar^2 B^2 + m^2\omega^2(2\hbar^2 + 2\hbar\sqrt{\hbar^2 - \hbar B\theta} - \hbar B\theta)}{m(2\hbar^2 + 2\hbar\sqrt{\hbar^2 - \hbar B\theta} - \hbar B\theta)}} \nonumber
\\
& \hspace{5mm} \sqrt{\frac{2\hbar^2 + 2\hbar\sqrt{\hbar^2 - \hbar B\theta} - \hbar B\theta + m^2\omega^2\theta^2}{4m\hbar^2}} \nonumber
\\
& \hspace{5mm} - m_l\hbar\frac{\hbar B + m^2\omega^2\theta}{2\hbar m},
\end{split}
\end{equation}
\end{comment}
where $n_r \in \mathbb{N}$ and $m_l \in \mathbb{Z}$ are radial and angular momentum quantum numbers respectively. 
%$\mathbb{N}$ and $\mathbb{Z}$ are the sets of natural numbers and integers respectively. 
On the other hand, the expression of the wavefunctions is 
\begin{equation} \label{eq:2.19}
\Psi_{n_r,m_l}(r,\varphi) = \frac{1}{\sqrt{2 \pi r}} R_{n_r,m_l}(r) e^{im_l\varphi},
\end{equation}
in which $R_{n_r,l}(r)$ is the radial wavefunction given as follows
\begin{equation}
R_{n_r,m_l}(r) = \left(\frac{2M\Omega}{\hbar}\right)^\frac{1}{2} \sqrt{\frac{n_r!}{(n_r + |m_l|)!}}\sqrt{r} \left(\frac{M\Omega}{\hbar}r^2\right)^\frac{|m_l|}{2} 
\exp \left(-\frac{M\Omega}{2\hbar}r^2\right) L_{n_r}^{|m_l|} \left(\frac{M\Omega}{\hbar}r^2\right),
\end{equation}
and $L_{n_r}^{m_l}$ is the Laguerre polynomials \cite{source3,source5}. 
Realize that there is a condition to be satisfied to the solution of the eigenvalue problem before it can really be applied to a physical system. In \eqref{eq:2.11}, the expression denotes the effective mass of a particle in the oscillator potential which is real and greater than 0. Hence,
\begin{equation}
\hbar^2 - \hbar B\theta \geqslant 0.
\end{equation}
Since we can control the magnitude of the magnetic field, it has to be non-negative and real i.e, $B \geqslant 0$. For the noncommutativity parameter, it is also non-negative and real. However, we are not interested to analyze the situation at $\theta = 0$ since we are discussing the NCQM model. Hence,
\begin{equation} \label{eq:2.22}
0 \leqslant B\theta \leqslant \hbar.
\end{equation}
Due to the above constraint, we will analyze further the three possible cases in greater details in the upcoming sections i.e, 
%$ B\theta = 0 $, $ B\theta = \hbar $ and $ 0 < B\theta < \hbar $.
\begin{align}
    \text{Case I:}& \quad B\theta = 0, \nonumber
    \\
    \text{Case II:}& \quad B\theta = \hbar, \nonumber
    \\
    \text{Case III:}& \quad 0 < B\theta < \hbar.
\end{align}

\section{Without magnetic field}

\subsection{Case I: 
$B\theta = 0$}

In the absence of magnetic field, the solution of the eigenvalue problem in \eqref{eq:2.16} and \eqref{eq:2.19} can be simplified as follows
\begin{equation} \label{eq:3.1}
\begin{split}
%E_{n_r,m_l} & = \left(2n_r + |m_l| + 1\right)\hbar\Omega - m_l\hbar\gamma,
%\\
%& 
E_{n_r,m_l}
= \left(2n_r + |m_l| + 1\right)\hbar\sqrt{\omega^2\left(1 + \frac{m^2 \omega^2 \theta^2}{4\hbar^2}\right)} - m_l\hbar\frac{\vartheta}{2\hbar}m\omega^2.
\end{split}
\end{equation}
\begin{align} \label{eq:3.2}
\begin{aligned}	
\Psi^{\theta}_{n_r,m_l}(r,\varphi) = & \frac{1}{\sqrt{2 \pi}}\left(\frac{4m\omega}{\sqrt{4\hbar^2 + m^2 \omega^2 \theta^2}}\right)^\frac{1}{2}  \sqrt{\frac{n_r!}{(n_r + |m_l|)!}} \left(\frac{2m\omega}{\sqrt{4\hbar^2 + m^2 \omega^2 \theta^2}}r^2\right)^\frac{|m_l|}{2} 
\\
& \exp \left(-\frac{m\omega}{\sqrt{4\hbar^2 + m^2 \omega^2 \theta^2}}r^2\right)
 L_{n_r}^{|m_l|} \left(\frac{2m\omega}{\sqrt{4\hbar^2 + m^2 \omega^2 \theta^2 }}r^2\right) e^{im_l\varphi}.
\end{aligned}
\end{align}
A close inspection reveals that the eigenvalues and eigenstates shown in \eqref{eq:3.1} and \eqref{eq:3.2} are quite familiar in some literature (e.g \cite{Jellal2021,source3}) as it is actually the solution of the eigenvalue problem involving noncommutative planar isotropic harmonic oscillator if the coordinates transformation used is the generalized Bopp shift or Seiberg-Witten map. The energy eigenvalues of the system for the first few lower quantum number pairs are shown in Table \ref{tab:1}.

\begin{table}[H]
	\centering
		\caption{\sf Energy eigenvalues  of the first few ground and excited states designated by the quantum number pairs $(n_r,m_l)$ in terms of $\Omega$ and $\gamma$.}
	\label{tab:1}
		\begin{tabular}{|c|c|c|c|c|c|}
			\hline
		    $(n_r,0)$ & Energy & $(0,m_l)$ & Energy & $(n_r,m_l)$ & Energy \\
			\hline
			$(1,0)$ & $3\hbar\Omega$ & $(0,1)$ & $2\hbar\Omega - \hbar\gamma$ & $(1,-1)$ & $4\hbar\Omega + \hbar\gamma$ \\ 
			
			$(2,0)$  & $5\hbar\Omega$ & $(0,2)$ & $3\hbar\Omega - 2\hbar\gamma$ & $(1,-2)$ & $5\hbar\Omega + 2\hbar\gamma$ \\
			
			$(3,0)$  & $7\hbar\Omega$ & $(0,3)$  & $4\hbar\Omega - 3\hbar\gamma$ & $(2,-1)$ & $6\hbar\Omega + \hbar\gamma$ \\
			
			&& $(0,-1)$ & $2\hbar\Omega + \hbar\gamma$ & $(1,1)$ & $4\hbar\Omega - \hbar\gamma$ \\
			
			&& $(0,-2)$  & $3\hbar\Omega + 2\hbar\gamma$ & $(1,2)$ & $5\hbar\Omega - 2\hbar\gamma$ \\
			
			&& $(0,-3)$  & $4\hbar\Omega + 3\hbar\gamma$ & $(2,1)$ & $6\hbar\Omega - \hbar\gamma$ \\
			\hline
		\end{tabular} 
\end{table}
As can be seen, it is not immediately obvious to infer if there is any hidden pattern in the distribution of energies. Hence, we will express $\gamma$ in terms of $\Omega$ in \eqref{eq:3.1} as this step is crucial in simplifying the analysis later on
\begin{align} \label{eq:3.5}
    \gamma = \sqrt{1 - \frac{\omega^2}{\Omega^2}} = \kappa\Omega,
\end{align}
where $0 < \kappa < 1$ is a direct consequence since the frequency, $\omega$ is nonzero. We can then rewrite the energies as tabulated in Table \ref{tab:2}.

\begin{figure}[H]
    \centering
    \includegraphics[scale=0.6]{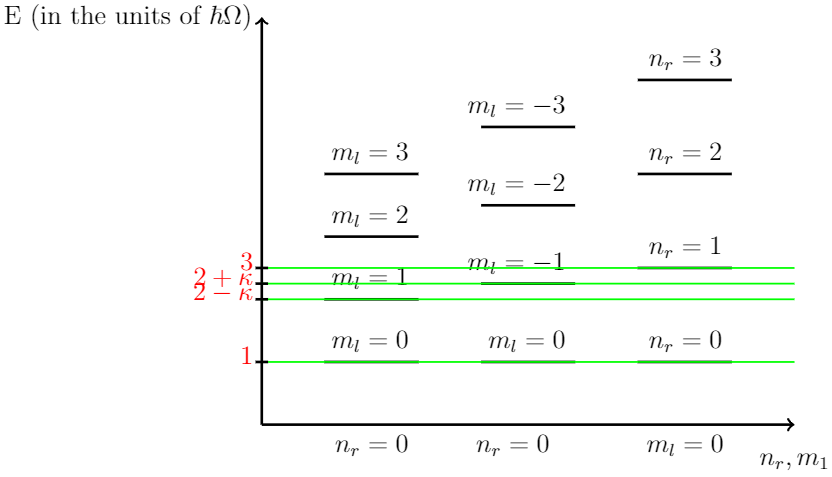}
    \caption{\sf (color online) Energy level diagram for the first few ground and excited states designated by the quantum number pairs $(n_r,0)$ and $(0,m_l)$.}
    \label{fig:3}
\end{figure}

If one of the two quantum numbers is zero, it will result in the energy spectrum which is equidistant from one another as can be observed more clearly in Figure \ref{fig:3}. Clearly, we can notice that the effect of the individual quantum numbers $n_r$ and $m_l$ differs ever so slightly in the minimal discrete energy step or quanta of energy, $\delta E$ depending on $\kappa$. The sign of $m_l$ also plays a role in affecting the gap of $\delta E$.
\begin{align}
	& \delta E_{m_l>0} < \delta E_{m_l<0} < \delta E_{n_r}, \nonumber
	\\
	&    (1 - \kappa)\hbar\Omega < (1 + \kappa) < 2\hbar\Omega.
\end{align}
\begin{table}[H]
	\centering
		\caption{\sf Energy eigenvalues  of the first few ground and excited states designated by the quantum number pairs $(n_r,m_l)$ in terms of $\Omega$.}
	\label{tab:2}
		\begin{tabular}{|c|c|c|c|c|c|}
			\hline
		    $(n_r,0)$ & Energy & $(0,m_l)$ & Energy & $(n_r,m_l)$ & Energy \\
			\hline
			$(1,0)$ & $3\hbar\Omega$ & $(0,1)$ & $ (2 - \kappa)\hbar\Omega$ & $(1,-1)$ & $(4+\kappa)\hbar\Omega$ \\ 
			
			$(2,0)$  & $5\hbar\Omega$ & $(0,2)$ & $ (3 - 2\kappa) \hbar\Omega$ & $(1,-2)$ & $(5+2\kappa)\hbar\Omega$ \\
			
			$(3,0)$  & $7\hbar\Omega$ & $(0,3)$  & $ (4 - 3\kappa) \hbar\Omega$ & $(2,-1)$ & $(6+\kappa)\hbar\Omega$ \\ 
			
			&& $(0,-1)$ & $ (2 + \kappa) \hbar\Omega$ & $(1,1)$ & $(4-\kappa)\hbar\Omega$ \\
			
			&& $(0,-2)$  & $ (3 + 2\kappa) \hbar\Omega$ & $(1,2)$ & $(5-2\kappa)\hbar\Omega$ \\
			
			&& $(0,-3)$  & $ (4 + 3\kappa) \hbar\Omega$ & $(2,1)$ & $(6-\kappa)\hbar\Omega$ \\
			\hline
		\end{tabular}
\end{table}

However, what really matters is the joint effect of both quantum numbers that has to be determined to see if the system has any degeneracy. At least, any two successive degenerate energy levels can always be determined if there exists, such as
\begin{align}
E_{n_r;1,m_l;1}  = E_{n_r;2,m_l;2},	
\end{align}
otherwise we have
\begin{align} 
	\begin{aligned}
  &  \left(2n_{r;1} + |m_{l;1}| + 1\right)\hbar\Omega - m_{l;1}\hbar \kappa\Omega  = \left(2n_{r;2} + |m_{l;2}| + 1\right)\hbar\Omega - m_{l;2}\hbar \kappa\Omega,
    \\
  &  \kappa = \frac{2(n_{r;2} - n_{r;1}) + (|m_{l;2}| - |m_{l;1}|)}{m_{l;2} - m_{l;1}}.\label{eq:3.8}
\end{aligned}
\end{align}
\begin{comment}
    \left(2n_{r;1} + |m_{l;1}| + 1\right) - m_{l;1} \kappa & = \left(2n_{r;2} + |m_{l;2}| + 1\right) - m_{l;2} \kappa, \nonumber
    \\
    - m_{l;1} \kappa + m_{l;2} \kappa & = \left(2n_{r;2} + |m_{l;2}| + 1\right) - \left(2n_{r;1} + |m_{l;1}| + 1\right), \nonumber
\end{comment}
Hence, we can compare $\kappa$ from \eqref{eq:3.5} and \eqref{eq:3.8} to get
\begin{align} \label{eq:3.7}
    \kappa = \frac{2(n_{r;2} - n_{r;1}) + (|m_{l;2}| - |m_{l;1}|)}{m_{l;2} - m_{l;1}}   = \sqrt{1 - \frac{4\hbar^2}{4\hbar^2 + m^2 \omega^2 \theta^2}}.
\end{align}
\begin{comment}
    & = \sqrt{1 - \omega^2\left[\omega^2\left(1 + \frac{m^2 \omega^2 \theta^2}{4\hbar^2}\right)\right]^{-1}} 
    \\
    & = \sqrt{1 - \left(1 + \frac{m^2 \omega^2 \theta^2}{4\hbar^2}\right)^{-1}},
\end{comment}
As long as the above equation is satisfied, the degenerate energy levels can be found. The left hand side of the equation above is a rational fraction. Then, the corresponding right hand side has to be a rational fraction as well. As a result, it is natural to eliminate the $\hbar^2$-term. That is only possible if the product $m^2\omega^2\theta^2$ is of the order of a positive rational number multiple of $\hbar^2$ i.e, $m^2\omega^2\theta^2 = c^2\hbar^2$. Hence, we can write
\begin{align} \label{eq:3.9}
    \kappa = \sqrt{\frac{c^2}{4 + c^2}}. 
\end{align}
Then, by using the method of induction (where the relevance of the square term in $c^2$ is in simplifying the method), the sequence of possible values of $c^2$ can be determined as follows
\begin{align}
    c_{n,k}^2 =& \frac{k^2}{n(n+k)},
\end{align}
\begin{comment}
\begin{align}
    \frac{1}{2},\frac{1}{6},\frac{1}{12},\frac{1}{20},\frac{1}{30},...&,\frac{1}{n(n+1)}, \nonumber
    \\
    \frac{2^2}{3},\frac{2^2}{8},\frac{2^2}{15},\frac{2^2}{24},\frac{2^2}{35},...&,\frac{2^2}{n(n+2)}, \nonumber
    \\
    \frac{3^2}{4},\frac{3^2}{10},\frac{3^2}{18},\frac{3^2}{28},\frac{3^2}{40},...&,\frac{3^2}{n(n+3)}, \nonumber
    \\
    & \vdots \nonumber
\end{align}
\end{comment}
where we will attach the subscript i.e, $c_{n,k}$ in the sequel. To check the validity of the above proposed sequence
\begin{align} \label{eq:3.12}
    \sqrt{\frac{c_{n,k}^2}{4+c_{n,k}^2}} = \frac{k}{2n + k},
\end{align}
\begin{comment}
    & = \sqrt{\frac{\frac{k^2}{n(n+k)}}{4 + \frac{k^2}{n(n+k)}}}
    \\
    & = \sqrt{\frac{\frac{k^2}{n(n+k)}}{\frac{4n(n+k)+k^2}{n(n+k)}}}, \nonumber
    \\
    & = \sqrt{\frac{k^2}{4n(n+k)+k^2}}, \nonumber
    \\
    & = \sqrt{\frac{k^2}{4n^2 + 4nk + k^2}}, \nonumber
    \\
    & = \sqrt{\frac{k^2}{(2n + k)^2}}, \nonumber
\end{comment}
which is indeed rational. To sum up, the degenerate energy levels of the system in Case I i.e, in the absence of magnetic field where the system is similar to the noncommutative planar harmonic oscillator under minimal coupling prescription can always be found and is very unique to the particular system provided that the noncommutativity can be described in terms of mass and frequency of the particle that is
\begin{align} \label{eq:3.10}
    \theta_d = \Bigg\{\left(\frac{\hbar}{m\omega}\right)c_{n,k} \Bigg| c_{n,k} = \frac{k}{\sqrt{n(n+k)}}, n,k \in \mathbb{Z} > 0 \Bigg\}.
\end{align}
Notice that the above result is also dimensionally consistent since, the noncommutativity is in the units of $m^2$ whereas mass, angular frequency and reduced Planck constant are in the units of kg, $\text{s}^{-1}$ and $\text{m}^2 \cdot \text{kg} \cdot \text{s}^{-1}$. The factor $c_{n,k}$ is dimensionless. We also want to note that similar study for Case 1 has been done in \cite{source1} with a different approach i.e, through algebraic method. The difference is, they describe the degenerate noncommutativity in terms of two relatively prime numbers whereas we describe it in terms of two positive integers.

We can also say that everytime the energy is of the form
\begin{align}
    E = (2n_r + |m_l| + 1)\hbar\Omega - \frac{k}{2n+k} m_l \hbar \Omega,
\end{align}
we can expect to have degenerate energy levels for any pair of quantum numbers. The uniqueness of degeneracy comes from the fact that apart from the parameter $c_{n,k}$, the noncommutativity, $\theta$ in $\eqref{eq:3.10}$ is actually a function of mass and frequency of the particle. These parameters influence the resulting quantum number pairs to be obtained to produce degenerate states.
\begin{comment}
 Let us substitute \eqref{eq:3.10} into $\Omega$ and observe what happens
\begin{align}
    \Omega & = \sqrt{\omega^2\left(1 + \frac{m^2\omega^2\left(\frac{\hbar}{m\omega}\right)^2(c_{n,k})^2}{4\hbar^2}\right)} \nonumber
    \\
    & = \sqrt{\omega^2\left(1 + \frac{k^2}{4n(n+k)}\right)}
\end{align}
The implication of the above result is clear. It means that while mass is still needed to find the degenerate $\theta$, finding degenerate energies do not require mass.
\end{comment}

\begin{comment}
Recall that each of the angular momentum quantum number can either be positive and negative integers. To see the degeneracy pattern, we will let the quantum numbers to be positive to ease the calculation. The negative ones will be compensated with a negative sign in front of their corresponding symbols. 
\end{comment}
By equating the left-hand side of \eqref{eq:3.7} and the right-hand side of \eqref{eq:3.12}, the three successive degenerate energy levels can be found. When the angular momentum quantum numbers for any two degenerate levels are both positive
\begin{comment}
\begin{align}
    \frac{2(n_{r;2} - n_{r;1}) + (m_{l;2} - m_{l;1})}{m_{l;2} - m_{l;1}} = \frac{k}{2n+k} = \frac{-k}{-(2n+k)}, \nonumber
    \\
    2(n_{r;2} - n_{r;1}) + 2n + k = k \quad \mbox{or} \quad 2(n_{r;2} - n_{r;1}) - (2n + k) = -k.
\end{align}
We assume that both the left and right-hand side of the equation are in their simplest form. Then
\end{comment}
\begin{align} \label{eq:3.15}
    E_{n_{r;1},m_{l;1}} = E_{n_{r;1} - n,m_{l;1} + 2n+k} = E_{n_{r;1} + n,m_{l;1} - (2n+k)}.
\end{align}
If however, we are focusing on both angular momentum quantum numbers to be negative, then
\begin{comment}
\begin{align}
    \frac{2(n_{r;2} - n_{r;1}) + (|-m_{l;2}| - |-m_{l;1}|)}{-m_{l;2} - (-m_{l;1})} & = \frac{k}{2n+k} = \frac{-k}{-(2n+k)}, \nonumber
    \\
    \frac{2(n_{r;2} - n_{r;1}) - (-m_{l;2} + m_{l;1})}{-m_{l;2} + m_{l;1}} & = \frac{k}{2n+k} = \frac{-k}{-(2n+k)}, \nonumber
    \\
    2(n_{r;2} - n_{r;1}) - (2n+k) = k \quad &\mbox{or} \quad 2(n_{r;2} - n_{r;1}) + 2n+k = -k.
\end{align}
\end{comment}
\begin{align} \label{eq:3.17}
    E_{n_{r;1},-m_{l;1}} = E_{n_{r;1} + n+k,-m_{l;1} + 2n+k} = E_{n_{r;1} - (n+k),-m_{l;1} - (2n+k)}.
\end{align}
We will not discuss the situation when either one of the angular momentum quantum number is negative as we can just compare the energy levels of these two formulas to see their equivalence. Hence, any successive degenerate energy levels for a particular $\theta$ and hence excited states of the same energy can always be found when \eqref{eq:3.15} and \eqref{eq:3.17} are satisfied. Otherwise, the system is non-degenerate. 

\subsection{Example}

As an instructive example, let us consider the particle of interest to be an electron where its mass is
\begin{align}
    m = 9.109 \times 10^{-31} \,\text{kg}.
\end{align}
As for the angular frequency, it really depends on numerous factors in a particular system and is rather difficult to calculate accurately, but are generally of a similar order to those found in common household and industrial springs. Typical values lie in the range from $100 \text{N} \text{m}^{-1}$ to $1000 \text{N} \text{m}^{-1}$ \cite{source10}. The relationship between spring constant, $k$ and angular frequency, $\omega$ is given as follows
\begin{equation}
\omega = \sqrt{\frac{k}{m}},
\end{equation}
Hence, the acceptable domain of angular frequency will be somewhere between $1.048 \times 10^{16}$ and $3.313 \times 10^{16}$. If an electron is accelerated through a $10V$ electrostatic potential, the frequency in the non-relativistic limit follows \cite{Zurcher2016}
\begin{align}
    \omega = 1.518 \times 10^{16} \text{s}^{-1}.
\end{align}
Then, for simplicity, we will let $n=k$. At $n=k$, it does not matter what the actual value of this constant is as it will always produce $c_{n,k} = \mathlarger{\frac{1}{\sqrt{2}}}$. Hence, the noncommutativity parameter will be
\begin{align}
    \theta_d = \frac{1}{\sqrt{2}}\left(\frac{\hbar}{m\omega}\right) = 5.395 \times 10^{-21} \text{m}^2,
\end{align}
and therefore we obtain
\begin{align} \label{eq:3.9}
    \kappa & = \frac{1}{3}.
\end{align}
\begin{comment}
Then, for the successive degenerate energy levels, we will let $n=k=1$ as the choice does matter and hence
\begin{align}
    E_{n_{r;1},m_{l;1}} = E_{n_{r;1} - 1,m_{l;1} + 3} = E_{n_{r;1} + 1,m_{l;1} - 3}, \nonumber
    \\
    E_{n_{r;1},-m_{l;1}} = E_{n_{r;1} + 2,-m_{l;1} + 3} = E_{n_{r;1} - 2,-m_{l;1} - 3}.
\end{align}
\end{comment}
Table \ref{tab:7} and \ref{tab:8} below show the energy eigenvalues of the case
%Case I i.e, 
$B\theta = 0$ at $\theta = 5.395 \times 10^{-21} \text{m}^2$ for different quantum number pairs and hence excited states.
\begin{table}[H]
	\centering
		\caption{\sf (color online) Ordered pair of quantum numbers and its corresponding energy for $m_l \geq 0$ at $B\theta = 0$ and $\kappa = \frac{1}{3}$.}
		\label{tab:7}
		\begin{tabular}{|p{5cm}|p{7cm}|}
			\hline
			Energy (in units of $\hbar\Omega$)& $(n_r,m_l)$\\
			\hline
			$25\times\frac{1}{3}$ & $(3,2),(2,5),(1,8),(0,11)$ \\
			$23\times\frac{1}{3}$ & $(3,1),(2,4),(1,7),(0,10)$ \\
			$21\times\frac{1}{3} = 7$ & $\color{blue} (3,0),(2,3),(1,6),(0,9)$ \\
			$19\times\frac{1}{3}$ & $(2,2),(1,5),(0,8)$ \\
			$17\times\frac{1}{3}$ & $(2,1),(1,4),(0,7)$ \\
			$15\times\frac{1}{3} = 5$ & $\color{blue} (2,0),(1,3),(0,6)$ \\
			$13\times\frac{1}{3}$ & $(1,2),(0,5)$ \\
			$11\times\frac{1}{3}$ & $(1,1),(0,4)$ \\
			$9\times\frac{1}{3} = 3$ & $\color{blue} (1,0),(0,3)$ \\
			$7\times\frac{1}{3}$ & $(0,2)$ \\ 
			$5\times\frac{1}{3}$ & $(0,1)$ \\ 
			$3\times\frac{1}{3} = 1$ & $\color{blue} (0,0)$ \\ 
			\hline
		\end{tabular} 
\end{table}
\noindent 
We are highlighting the table to show that for every $3$ times of the minimal discrete energy step, $3\times\delta E$, there will be a single additional degenerate state in the subsequent levels. This pattern persists for higher-order states as well for all $m_l \geq 0$.

\begin{table}[H]
	\centering
		\caption{\sf (color online) Ordered pair of quantum numbers and its corresponding energy for $m_l < 0$ at $B\theta = 0$ and $\kappa = \frac{1}{3}$.}
		\label{tab:8}
		\begin{tabular}{|p{5cm}|p{7cm}|}
			\hline
			Energy (in units of $\hbar\Omega$)& $(n_r,m_l)$\\
			\hline
			$53\times\frac{1}{3}$ & $(1,-11),(3,-8),(5,-5),(7,-2)$ \\
			$51\times\frac{1}{3} = 17$ & $\color{blue} (0,-12),(2,-9),(4,-6),(6,-3)$ \\
			$49\times\frac{1}{3}$ & $(1,-10),(3,-7),(5,-4),(7,-1)$ \\
			$47\times\frac{1}{3}$ & $\color{blue} (0,-11),(2,-8),(4,-5),(6,-2)$ \\
			$45\times\frac{1}{3} = 15$ & $(1,-9),(3,-6),(5,-3)$ \\
			$43\times\frac{1}{3}$ & $\color{blue} (0,-10),(2,-7),(4,-4),(6,-1)$ \\
			$41\times\frac{1}{3}$ & $(1,-8),(3,-5),(5,-2)$ \\
			$39\times\frac{1}{3} = 13$ & $\color{blue} (0,-9),(2,-6),(4,-3)$ \\
			$37\times\frac{1}{3}$ & $(1,-7),(3,-4),(5,-1)$ \\
			$35\times\frac{1}{3}$ & $\color{blue} (0,-8),(2,-5),(4,-2)$ \\
			$33\times\frac{1}{3} = 11$ & $(1,-6),(3,-3)$ \\
			$31\times\frac{1}{3}$ & $\color{blue} (0,-7),(2,-4),(4,-1)$ \\
			$29\times\frac{1}{3}$ & $(1,-5),(3,-2)$ \\
			$27\times\frac{1}{3} = 9$ & $\color{blue} (0,-6),(2,-3)$ \\
			$25 \times \frac{1}{3}$ & $(1,-4),(3,-1)$ \\
			$23 \times \frac{1}{3}$ & $\color{blue} (0,-5),(2,-2)$ \\
			$21\times\frac{1}{3} = 7$ & $(1,-3)$ \\
			$19 \times \frac{1}{3}$ & $\color{blue} (0,-4),(2,-1)$ \\
			$17 \times \frac{1}{3}$ & $(1,-2)$ \\
			$15\times\frac{1}{3} = 5$ & $\color{blue} (0,-3)$ \\
			$13 \times \frac{1}{3}$ & $(1,-1)$ \\
			$11 \times \frac{1}{3}$ & $\color{blue} (0,-2)$ \\
			$7 \times \frac{1}{3}$ & $\color{blue} (0,-1)$ \\
			\hline
		\end{tabular} 
\end{table}

\begin{figure}[H]
    \centering
    \includegraphics[scale=0.55]{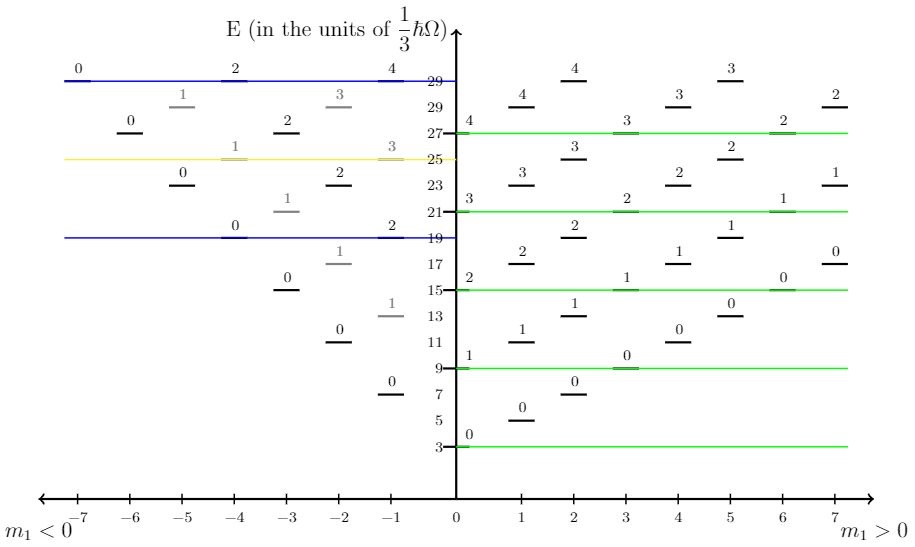}
    \caption{\sf (color online) Energy level diagram of the case  $B\theta = 0$ at $\kappa = \frac{1}{3}$.}
    \label{fig:2}
\end{figure}

For the remaining case of $m_l < 0$, the degeneracy pattern is not apparent at first glance. However, when we highlight the table to separate the energy levels consisting of even and odd $n_r$ states and treat them individually, we start to notice that the behaviour is more or less similar to the ones that we have earlier when $m_l \geq 0$. Figure \ref{fig:2} displays the side-by-side comparison of the asymmetric distribution of energy eigenvalues of positive and negative $m_l$ states.
Every integer on the black lines (energy levels) represents radial quantum number, $n_r$.
Each of the green line in Figure \ref{fig:2} signifies the level at which we start to have a single additional degenerate state compared to the previous line. The region in between them should be occupied by a similar number of degenerate states. The same behaviour is true for the negative $m_l$ domain though, to separate the even and odd $n_r$ states, we use blue and yellow lines respectively.

\section{With magnetic field}

In the presence of magnetic field, based on \eqref{eq:2.22}, the noncommutativity should be restricted in such a way that the product $B\theta$ is between 0 and $\hbar$. In this section, we will explore the remaining cases i.e, $B\theta = \hbar$ and $0 < B\theta < \hbar$ respectively.

\subsection{Case I: $B\theta = \hbar$}

The solution of the eigenvalue problem in \eqref{eq:2.16} and \eqref{eq:2.19} for the extreme end, $B\theta = \hbar$ can be simplified as follows
\begin{equation}\label{eq:4.1}
\begin{split}
E_{n_r,m_l}  = \left(2n_r + |m_l| + 1\right)\hbar\sqrt{\left(\frac{B^2 + m^2\omega^2}{m}\right) \left(\frac{\hbar^2 + m^2\omega^2\theta^2}{4m\hbar^2}\right)} - m_l\hbar\frac{\hbar B + m^2\omega^2\theta}{2\hbar m},
\end{split}
\end{equation}
\begin{align}
	\begin{aligned}
    \Psi_{n_r,m_l}^{B,\theta}(r,\varphi) = & \frac{1}{\sqrt{2 \pi}}\left(\frac{2}{\hbar} \sqrt{\frac{4\hbar^2(B^2 + m^2\omega^2)}{\hbar^2 + m^2\omega^2\theta^2}} \right)^\frac{1}{2}  \sqrt{\frac{n_r!}{(n_r + |m_l|)!}} \left(\frac{1}{\hbar} \sqrt{\frac{4\hbar^2(B^2 + m^2\omega^2)}{\hbar^2 + m^2\omega^2\theta^2}} r^2\right)^\frac{|m_l|}{2} 
    \\
    & \exp \left(-\frac{1}{2\hbar} \sqrt{\frac{4\hbar^2(B^2 + m^2\omega^2)}{\hbar^2 + m^2\omega^2\theta^2}} r^2\right)
   L_{n_r}^{|m_l|} \left(\frac{1}{\hbar} \sqrt{\frac{4\hbar^2(B^2 + m^2\omega^2)}{\hbar^2 + m^2\omega^2\theta^2}} r^2\right)e^{im_l\varphi}.
\end{aligned}  
\end{align}     
A quick check while taking into account that $B\theta = \hbar$ in \eqref{eq:4.1} yields $\Omega^2 = \gamma^2$ which implies that
\begin{comment}
\begin{align}
    \Omega^2 & = \frac{\hbar^2 B^2 + B^2m^2\omega^2\theta^2 + \hbar^2m^2\omega^2 + m^4\omega^4\theta^2}{4m^2\hbar^2}, \nonumber
    \\
    & = \frac{\hbar^2 B^2 + 2\hbar^2m^2\omega^2 + m^4\omega^4\theta^2}{4m^2\hbar^2}, \nonumber
    \\
    & = \frac{B^2}{4m^2} + \frac{\omega^2}{2} + \frac{m^2\omega^4\theta^2}{4\hbar^2},
    \\
    \gamma^2 & = \frac{\hbar^2 B^2 + 2\hbar B m^2\omega^2\theta + m^4\omega^4\theta^2}{4\hbar^2 m^2}, \nonumber
    \\
    & = \frac{\hbar^2 B^2 + 2\hbar^2 m^2\omega^2 + m^4\omega^4\theta^2}{4\hbar^2 m^2}, \nonumber
    \\
    & = \frac{B^2}{4m^2} + \frac{\omega^2}{2} + \frac{m^2\omega^4\theta^2}{4\hbar^2}.
\end{align}
\end{comment}
\begin{align}
    \Omega = \gamma.
\end{align}
If we look at the energy eigenvalues closely when $\Omega=\gamma$, the degeneracy pattern is actually equivalent to those in Landau problem in symmetric gauge. The degeneracy of Landau problem in symmetric gauge is very well known and therefore, we will not provide any example. Every energy level of Landau problem in symmetric gauge is infinitely degenerate and hence, this applies to this case. We can then rewrite the energy eigenvalues as
\begin{align}
    (2n_r + |m_l| + 1 - m_l)\hbar\Omega. 
\end{align}
As refresher, the distribution of energies is shown in Figure \ref{fig:Landau} below. Unlike the first case, we can set $\theta$ to be of any value and we will still get the degenerate energy levels i.e,
\begin{align}
    \theta_d = \theta.
\end{align}

\begin{figure}[H]
    \centering
    \includegraphics[scale=0.55]{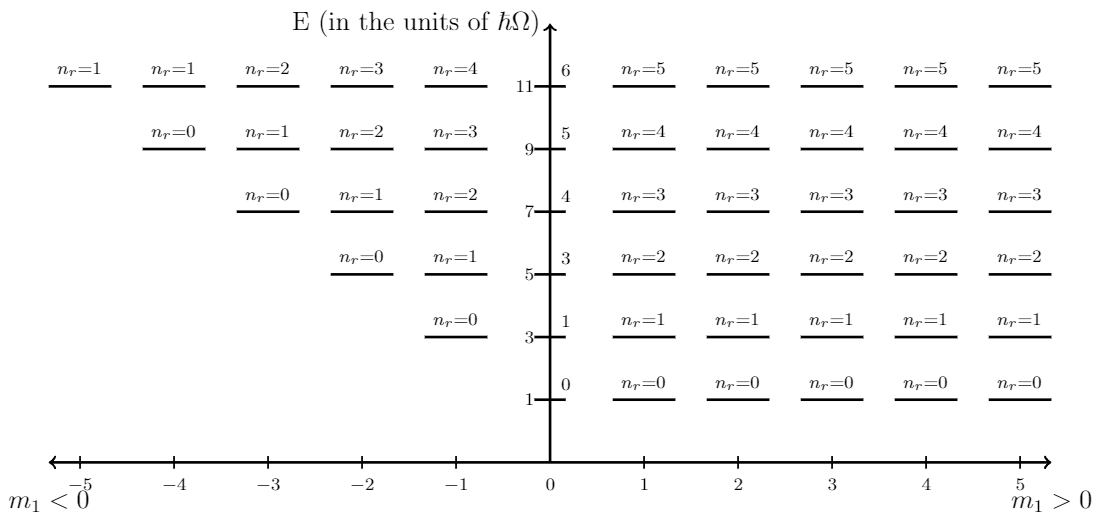}
    \caption{\sf Energy level diagram of the case  $B\theta = \hbar$ for $m_l < 0$.}
    \label{fig:Landau}
\end{figure}

\begin{comment}
\begin{figure}[H]
    \centering
    \fbox{\includegraphics[scale=0.7]{energylevel1.png}}
    \caption{Energy level diagram of Case 2 i.e, $B\theta = \hbar$ for $m_l < 0$}
\end{figure}

\begin{figure}[H]
    \centering
    \fbox{\includegraphics[scale=0.7]{energylevel2.png}}
    \caption{Energy level diagram of Case 2 i.e, $B\theta = \hbar$ for $m_l \geq 0$}
\end{figure}
\end{comment}

As can be seen, the infinite degeneracy in this setting is asymmetrical with respect to the sign of $m_l$ just like in the first case. The electron in response to the magnetic field and noncommutativity at $B\theta = \hbar$, irrespective of its individual value, appears to prefer a direction of $L_z$ requiring more energy to exist in the positive states than the negative states. In this noncommutative space, it costs energy for an electron circulating with positive angular momentum while it is not the case when circulating with negative angular momentum. \cite{edgestates}.

\subsection{Case III: $0 < B\theta < \hbar$}

%For the last case of the domain $0 < B\theta < \hbar$, both magnetic field and noncommutativity are present. Back in Section \ref{sec:2}, we recall that
%\begin{align}
%\begin{aligned}	
%&\Omega  = \sqrt{\frac{\hbar^2 B^2 + m^2\omega^2(2\hbar^2 + 2\hbar\sqrt{\hbar^2 - \hbar B\theta} - \hbar B\theta)}{m(2\hbar^2 + 2\hbar\sqrt{\hbar^2 - \hbar B\theta} - \hbar B\theta)}} \sqrt{\frac{2\hbar^2 + 2\hbar\sqrt{\hbar^2 - \hbar B\theta} - \hbar B\theta + m^2\omega^2\theta^2}{4m\hbar^2}}, \\
%&
%\gamma  = \frac{\hbar B + m^2\omega^2\theta}{2\hbar m}.
%\end{aligned}
%\end{align}
\begin{comment}
By squaring the expression for each variable 
\begin{align}
    \Omega^2 & = \left[\frac{\hbar^2 B^2 + m^2\omega^2(2\hbar^2 + 2\hbar\sqrt{\hbar^2 - \hbar B\theta} - \hbar B\theta)}{m(2\hbar^2 + 2\hbar\sqrt{\hbar^2 - \hbar B\theta} - \hbar B\theta)}\right] \left[\frac{2\hbar^2 + 2\hbar\sqrt{\hbar^2 - \hbar B\theta} - \hbar B\theta + m^2\omega^2\theta^2}{4m\hbar^2}\right], \nonumber
\end{align}
\begin{align}
    \gamma^2 & = \frac{\hbar^2 B^2 + 2\hbar B m^2\omega^2\theta + m^4\omega^4\theta^2}{4\hbar^2 m^2}.
\end{align}
\end{comment}
By squaring and rearranging the expressions \eqref{omega} and \eqref{gamma} %for each variable 
will enable us to spot the common and distinct terms
\begin{align}
\Omega^2 & = \frac{B^2}{4m^2} + \left[\frac{B^2\omega^2\theta^2}{4(2\hbar^2 + 2\hbar\sqrt{\hbar^2 - \hbar B \theta} - \hbar B \theta)} + \frac{\omega^2(2\hbar^2 + 2\hbar\sqrt{\hbar^2 - \hbar B \theta} - \hbar B \theta)}{4\hbar^2}\right] + \frac{m^2\omega^4\theta^2}{4\hbar^2}, 
\\
\gamma^2 & = \frac{B^2}{4m^2} + \left[\frac{\omega^2 B\theta}{2\hbar}\right] + \frac{m^2\omega^4\theta^2}{4\hbar^2},
\end{align}
Just like in Case I, after a few algebraic manipulation steps, we can then express $\gamma$ in terms of $\Omega$ as follows
\begin{align}
    \gamma^2 & = \xi\Omega^2,
\end{align}
and
it is straightforward to see that $\xi$ takes the form
\begin{align} \label{eq:4.9}
&\xi=\\
 &   \sqrt{1 - \frac{1}{\Omega^2} \left[\frac{\hbar^2\omega^2B^2\theta^2 + \omega^2(2\hbar^2 + 2\hbar\sqrt{\hbar^2 - \hbar B \theta} - \hbar B \theta)^2 - 2\hbar\omega^2 B\theta(2\hbar^2 + 2\hbar\sqrt{\hbar^2 - \hbar B \theta} - \hbar B \theta)}{4\hbar^2(2\hbar^2 + 2\hbar\sqrt{\hbar^2 - \hbar B \theta} - \hbar B \theta)}\right]}.\nonumber
\end{align}
A simple check reveals that the domain of $\xi$ is really the same as in $\kappa$ i.e, $0 < \xi < 1$. 
\begin{comment}
The effect of the individual quantum number when the other is zero on the distribution of energy is similar to the ones shown in the first case as displayed in Figure \ref{fig:2} but now in terms of $\xi$ instead of $\kappa$. Hence, the discussion will not be repeated here. Apart from that, the expression of $\xi$ is really the same as $\kappa$ in the form of the ratio of quantum numbers as in the left-hand side of \eqref{eq:3.7} and hence, it has to be a positive rational number for the degeneracy to occur.
\end{comment}
From this point, the discussion and arguments that lead to degeneracy are very much similar to Case I and hence will not be repeated here. In particular, we are referring to the effect of individual quantum numbers as portrayed in Figure \ref{fig:2} and $\xi$ is equivalent to $\kappa$ in terms of the ratio of quantum numbers as in \eqref{eq:3.8}. Later on, we will let the magnetic field and noncommutativity to be expressed as follows
\begin{align}
    B = fm\omega,\qquad
    \theta = g\frac{\hbar}{m\omega},
\end{align}
as they will make the expression of $\xi$ in \eqref{eq:4.9} to be rational. In addition, $\theta$ is also dimensionally consistent just like Case I. As for $B$, if we consider it as the magnetic field, it is supposed to be in the units of $\text{kg} \cdot \text{A}^{-1} \cdot \text{s}^{-2}$ but as mentioned in the very first section, $B$ is really the magnetic field multiplied by a unit charge then, it will be in the units of $\text{kg} \cdot \text{s}^{-1}$ since the unit of charge vanishes. Therefore, $B$ is also dimensionally consistent. To make the expression in \eqref{eq:4.9} more readable we will let
\begin{align}
    \mathfrak{L} & = 2\hbar^2 + 2\hbar\sqrt{\hbar^2 - \hbar B \theta} - \hbar B \theta = \hbar^2(2 + 2\sqrt{1 - fg} - fg).
\end{align}
Then
\begin{align}
    \xi & = \frac{2(n_{r;2} - n_{r;1}) + (|m_{l;2}| - |m_{l;1}|)}{m_{l;2} - m_{l;1}} = \sqrt{1 - \frac{\hbar^2m^2\omega^2 B^2\theta^2 + m^2\omega^2\mathfrak{L}^2 - 2\hbar m^2\omega^2 B\theta\mathfrak{L}}{\hbar^2 B^2\mathfrak{L} + \hbar^2 m^2 \omega^2 B^2 \theta^2 + m^2 \omega^2 \mathfrak{L}^2 + m^4 \omega^4 \theta^2 \mathfrak{L}}}. \nonumber
\end{align}
To produce degenerate energy levels, the right-hand side has to be equal to the left-hand side which is rational. By substituting the proposed definition of $B$ and $\theta$, the simplified form of $\xi$ is
\begin{comment}
\begin{align}
    \xi &= \sqrt{1 - \frac{f^2g^2\hbar^4m^2\omega^2 + m^2\omega^2\mathfrak{L}^2 - 2fg\hbar^2 m^2\omega^2\mathfrak{L}}{f^2\hbar^2 m^2\omega^2\mathfrak{L} + f^2g^2\hbar^4 m^2 \omega^2 + m^2 \omega^2 \mathfrak{L}^2 + g^2 \hbar^2 m^2 \omega^2  \mathfrak{L}}} \nonumber
    \\
   % &= \sqrt{1 - \frac{f^2g^2 + (2+2\sqrt{1-fg}-fg)^2 - 2fg(2+2\sqrt{1-fg}-fg)}{f^2g^2 + (2+2\sqrt{1-fg}-fg)^2 + (f^2 + g^2)(2+2\sqrt{1-fg}-fg)}}, \nonumber
    %\\
    & = \sqrt{1 - \frac{\left[(2+2\sqrt{1-fg}-fg) - fg\right]^2}{\left[(2+2\sqrt{1-fg}-fg) + f^2\right]\left[(2+2\sqrt{1-fg}-fg) + g^2\right]}}, \nonumber
\end{align}
\end{comment}
\begin{comment}
The value $0$ is not included in the inequality since that can only be true if $f = -g$. That cannot happen since both $f$ and $g$ are positive. Notice that we have eliminated all the related parameters which may make the expression in the right-hand side irrational
\end{comment} 
\begin{align} \label{eq:5.14}
    \xi &= \sqrt{\frac{(f+g)^2}{4 + (f-g)^2}}.
\end{align}
Now, the problem of finding degenerate states essentially reduces to the problem of finding the appropriate values or sequence of values of $f$ and $g$ such that \eqref{eq:5.14} is rational.

By the method of induction, we find that the difference $f-g$ has to be either
\begin{align}  \label{eq:4.201}
    f - g = \frac{4nk}{n^2 - k^2},
\end{align}
or 
 \begin{align}
     f - g = \frac{n^2 - k^2}{nk},
     \label{eq:4.20}
\end{align}
where $n$ and $k$ are positive integers with $f > g$, and with $f$ and $g$ coprime and not both odd. Apart from that, since $0 < B\theta < \hbar$ and $B\theta = fg\hbar$ then
\begin{align}
    fg < 1.
\end{align}
Usually, we will set the value of magnetic field at our discretion in \textit{experiment} and hence, we will attach the subscript to the controlling parameter $f_{exp}$. It is all the different possibilities of the value of noncommutativity that has to be determined to locate degenerate states where we will then denote as $g_{n,k;f}$ with
\begin{align}
    g_{n,k;f} = f_{exp} - \frac{4nk}{n^2 - k^2}, 
\end{align}
or again
 \begin{align}   
     g_{n,k;f} = f_{exp} - \frac{n^2 - k^2}{nk},
\end{align}
such that the domain is $0 < g_{n,k;f} < \frac{1}{f_{exp}}$. We also want to make it clear that individually, $f_{exp}$ and $g_{n,k;f}$ has to be a positive rational number in \eqref{eq:5.14} as to ensure that the numerator is rational and hence $\xi$ is rational. Then, the degenerate noncommutativity will be
\begin{align}
    \theta_d = \Bigg\{\left(\frac{\hbar}{m\omega}\right)g_{n,k;f} \Bigg| g_{n,k;f} = \bigg\{f_{exp} - \frac{4nk}{n^2 - k^2},f_{exp} - \frac{n^2 - k^2}{nk}\bigg\}, \text{gcd}(n,k) = 1, n,k \in \mathbb{Z} > 0 \Bigg\}.
\end{align}
where $f_{exp}$ and $g_{n,k;f}$ are dimensionless factors.

\subsection{Example}

\begin{comment}
It is tempting to use the cyclotron frequency as it is usually used in literature. It can be denoted as follows
\begin{align}
    \omega_c = \frac{B}{m},
\end{align}
and to get degenerate states, we have to set 
\begin{align}
    B = fm\omega.
\end{align}
where $f = 1$ is the only way to achieve that. Then,
\begin{align}
    \frac{|k^2 - 2n^2|}{\sqrt{[(k-n)^2 + n^2][(k+n)^2 + n^2]}} \nonumber
    \\
    \frac{|k^2 - 2n^2|}{\sqrt{k^4 + 4n^4}}
\end{align}
has to be rational. However, it cannot be realized as the expression under the radical cannot be a perfect square. This is because, if it is a perfect square, the the corresponding Diophantine equation has no solution as proved by Fermat by the method of infinite descent. We refer to \cite{Diophantine} for greater details
\end{comment}
In quantum Hall effect experiment, the typical value of magnetic field is
\begin{align}
    B = 12q \text{kg s}^{-1} = 1.922 \times 10^{-18} \text{kg s}^{-1} = 12 \text{kg A}^{-1}\text{s}^{-2},
\end{align}
where $q$ is the charge of particle of interest (in this case, an electron) and is the typical magnitude being used in quantum Hall effect experiments \cite{QHE1,QHE2}. Since we will again consider an electron in Case I with mass $9.109 \times 10^{-31} \text{kg}$ and angular frequency $\omega = 1.518 \times 10^{16} \text{s}^{-1}$. Hence, 
\begin{align} \label{eq:4.22}
    f_{exp} = \frac{12q}{m\omega} = 1.390 \times 10^{-4},
\end{align}
which is clearly irrational. From here alone, we can actually infer that in general, it is actually very difficult to finely tune the magnetic field in \textit{experimental} settings in just the right value to be able to \textit{observe} the degenerate states. Consequently, we will then take the magnetic field to be of lower value such that \eqref{eq:4.22} is rational
\begin{align}
    f_{exp} \approx 1 \times 10^{-4} = \frac{1}{10000},
\end{align}
where now, $B = 8.631q \text{kg s}^{-1}$. We will use \eqref{eq:4.20} and by letting $n = 2 \times 10^4 + 1$ and $k = 2 \times 10^4$ (which are coprime and not both odd), then 
\begin{align}
    g_{n,k;f} = \frac{1}{10000} - \frac{(2 \times 10^4 + 1)^2 - (2 \times 10^4)^2}{(2 \times 10^4 + 1)(2 \times 10^4)} = \frac{1}{400020000}.
\end{align}
This results in the following degenerate noncommutativity
\begin{align}
    \theta_d = g_{n,k;f}\frac{\hbar}{m\omega} = 1.907 \times 10^{-29} \text{m}^2.
\end{align}
Notice that
\begin{align}
    f_{exp} - g_{n,k;f} = \frac{40001}{400020000},
\end{align}
which does in fact satisfy equation  \eqref{eq:4.20}. Then, we have
\begin{align}
    \xi = \sqrt{\frac{(f_{exp}+g_{n,k;f})^2}{4 + (f_{exp}-g_{n,k;f})^2}} = \frac{40003}{800040001}.
\end{align}
We will not however be plotting the energy level diagram for the above value of $\xi$ since its numerator and denominator are very large. That means, it is expected that the very first lowest degenerate state has a relatively high energy to be plotted in a proper scale. Hence, again for \textit{experimental} settings, we need to use sufficiently very high magnetic field (e.g, in the range of thousands of Tesla) to \textit{detect} degenerate states more easily that will in turn give more acceptable value of $\xi$. Since $0 < B\theta < \hbar$, high $B$ will make the acceptable domain of $\theta$ to be even smaller.
 
\begin{comment}
\begin{table}[H]
\caption{Effect of noncommutativity on the ground level probability density at $\mathbf{B} = \frac{1}{8}\hbar$ with varying $\theta$ at $r = 5 \times 10^{-17}$}
\centering
\tabulinesep=\tabcolsep
\begin{tabu} to \textwidth {|X[4,c,m]|X[4,c,m]|X[4,c,m]|}
    \hline
    \includegraphics[width=\linewidth,keepaspectratio]{contour1.png} &   \includegraphics[width=\linewidth,keepaspectratio]{contour2.png} & 
    \includegraphics[width=\linewidth,keepaspectratio]{contour3.png}
    \\ \hline
    (a) $[\Psi_{0,0}(r,\varphi)]^2$ at $\mathbf{\theta} = 2\hbar$ & (b) $[\Psi_{0,0}(r,\varphi)]^2$ at $\mathbf{\theta} = 4\hbar$ & (c) $[\Psi_{0,0}(r,\varphi)]^2$ at $\mathbf{\theta} = 6\hbar$
    \\ \hline
\end{tabu}
\end{table}
\end{comment}

\section{Effects of $B$ and $\theta$ on probability densities}
% and effect of magnetic field and noncommutativity}

In this section we are going to display the probability distribution functions of the ground and excited states and also the effect of magnetic field and noncommutativity for each one of the cases. Despite all the constraints imposed in these three cases, we want to note that the energy eigenvalues and eigenstates can all be traced back to \eqref{eq:2.16} and \eqref{eq:2.19} where we only need to manipulate $\Omega$ and $\gamma$ without changing the mathematical structure of functions. Hence, it is expected that the upcoming plots of probability densities will not look much different from one another.
\begin{comment}
The reason we decide to put them into a single section here is the fact that the energy spectra and wavefunctions all originates from the very first section. The only thing that differentiate them would be the factors of the energy eigenvalues and eigenstates themselves due to the different constraints being imposed through the magnetic field and noncommutativity on the system. 
\end{comment}

We will plot the probability densities of the first 25 ground and excited states to observe the behaviour of the function. As for the effect of magnetic field and noncommutativity, we will only present the ground state i.e, $|\Psi_{0,0}|^2$ as the effect can naturally be extended to higher-order states. For all the cases, we will set the mass and frequency of the particle and applied magnetic field to be similar to those in the examples of degenerate energy spectra from the previous sections i.e,
\begin{align}
    m = 9.109 \times 10^{-31}\, \text{kg} \qquad \omega = 1.518 \times 10^{16}\, \text{s}^{-1} \qquad B = 12q \,\text{kg s}^{-1},
\end{align}

\begin{figure}[H]
    \centering
    \includegraphics[scale=0.45]{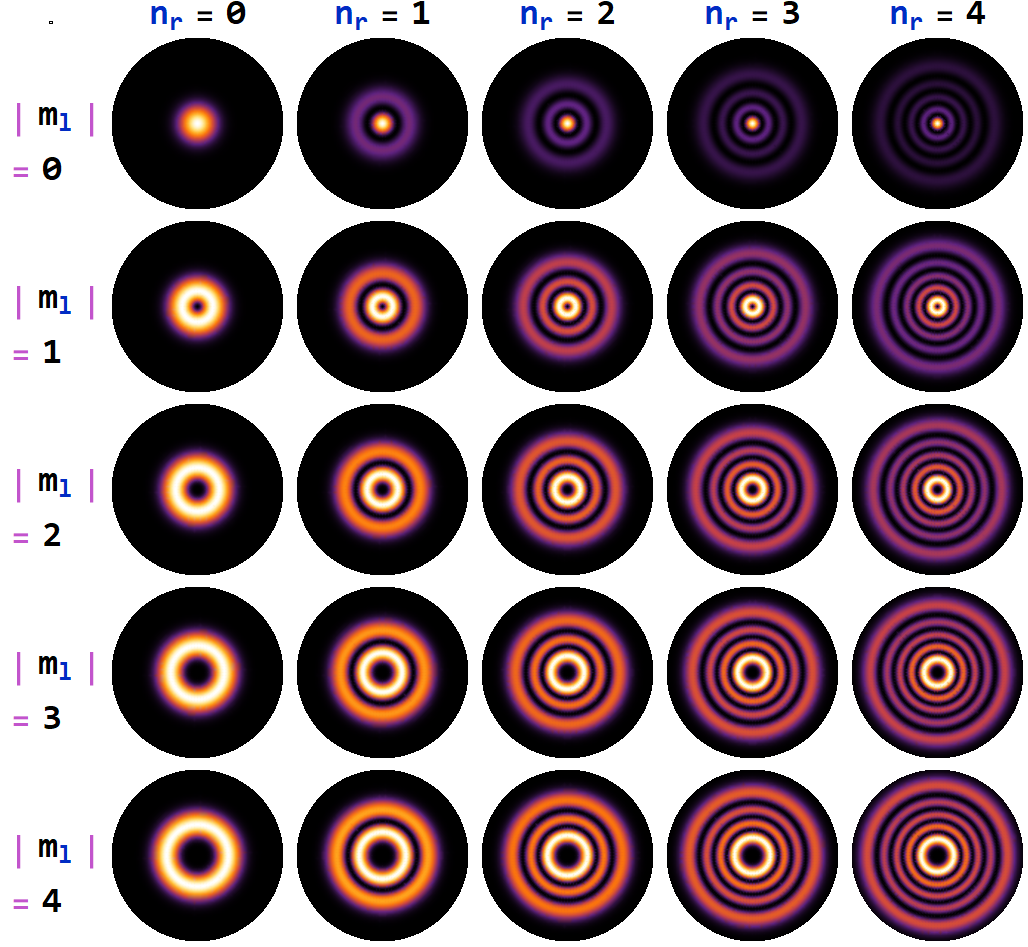}
    \caption{\sf (color online) Density plot of $|\Psi_{n_r,|m_l|}|^2$ at $B\theta = 0$ where $\theta = 5.395 \times 10^{-21}$ bounded by the radius $r = 5 \times 10^{-10}$ m.}
    \label{fig:4}
\end{figure}

\begin{figure}[H]
    \centering
    \includegraphics[scale=0.45]{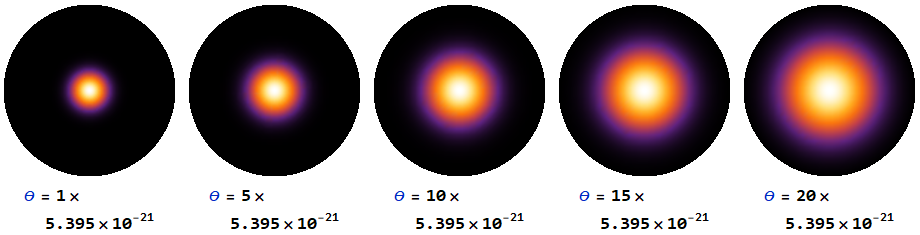}
    \caption{\sf (color online) Effect of $\theta$ on $|\Psi_{0,0}|^2$ at $B\theta = 0$ bounded by the radius $r = 5 \times 10^{-10}$ m.}
    \label{fig:5}
\end{figure}

\begin{figure}[H]
    \centering
    \includegraphics[scale=0.45]{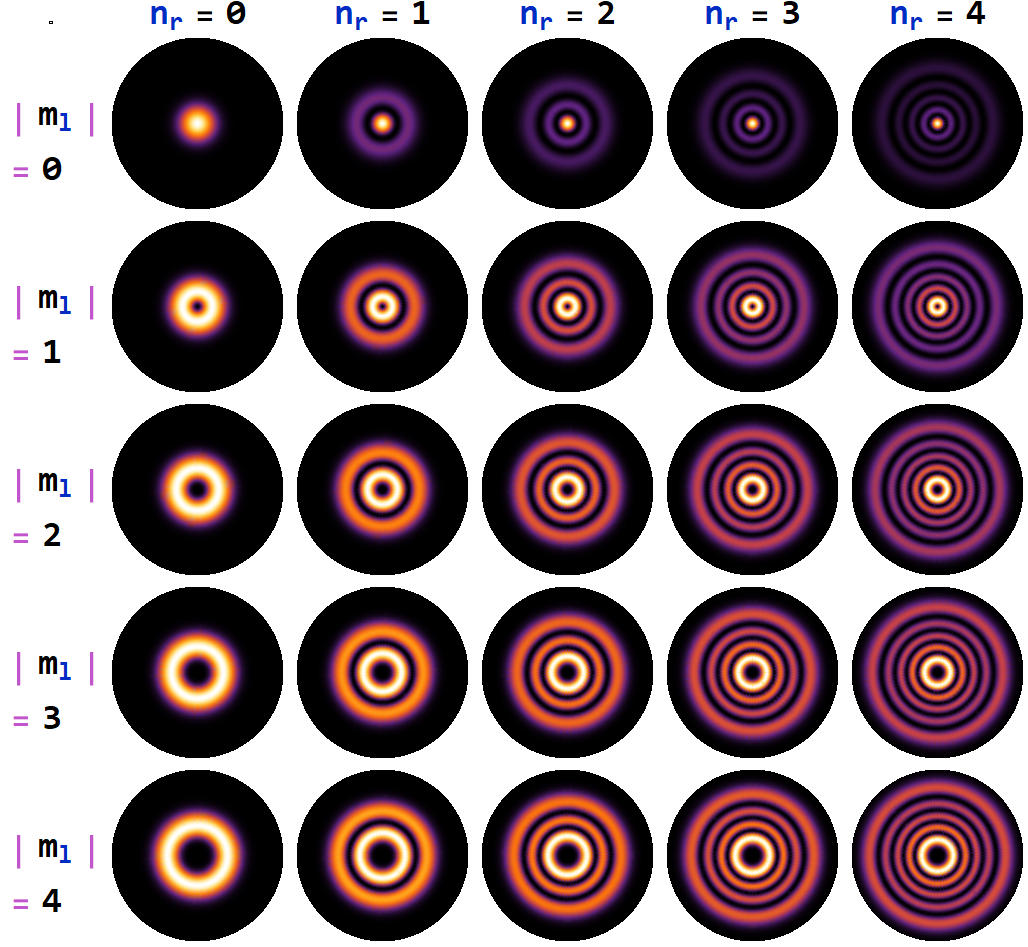}
    \caption{\sf (color online) Density plot of $|\Psi_{n_r,|m_l|}|^2$ at $B\theta = \hbar$ where $\theta = 5.488 \times 10^{-17}$ m$^2$ bounded by the radius $r = 3 \times 10^{-8}$ m.}
    \label{fig:6}
\end{figure}

\begin{figure}[H]
    \centering
    \includegraphics[scale=0.5]{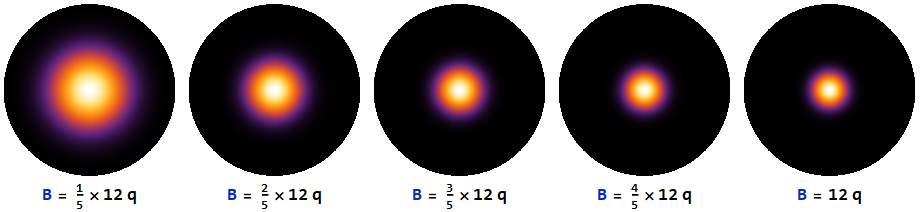}
    \caption{\sf (color online) Effect of $B$ on $|\Psi_{0,0}|^2$ at $B\theta = \hbar$ bounded by the radius $r = 3 \times 10^{-8}$ m.}
    \label{fig:7}
\end{figure}

\begin{figure}[H]
    \centering
    \includegraphics[scale=0.45]{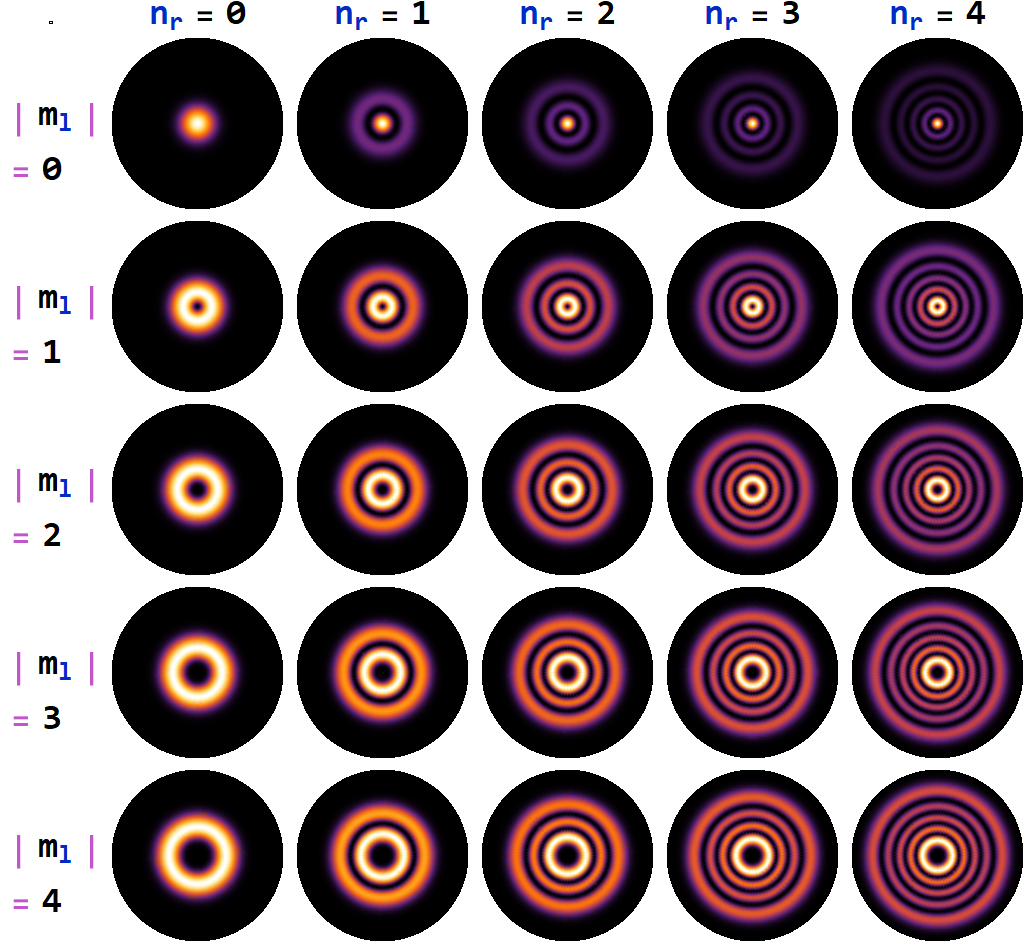}
    \caption{\sf (color online) Density plot of $|\Psi_{n_r,|m_l|}|^2$ at $0 < B\theta < \hbar$ where $\theta = 1.907 \times 10^{-29}$ m$^2$ bounded by the radius $r = 2 \times 10^{-9}$ m.}
    \label{fig:8}
\end{figure}

The density plots of the probability distribution functions for the ground state and the first few excited states are manifested in Figure \ref{fig:4}, Figure \ref{fig:6} and Figure \ref{fig:8} in Case I, Case II and Case III respectively. Accordingly, the effects of noncommutativity on the ground state probability density functions are shown in Figure \ref{fig:5}, Figure \ref{fig:7} and Figure \ref{fig:9}.

\subsection{Case I: $B\theta = 0$}

\subsection{Case II: $B\theta = \hbar$}

\subsection{Case III: $0 < B\theta < \hbar$}

\begin{figure}[H]
    \centering
    \includegraphics[scale=0.45]{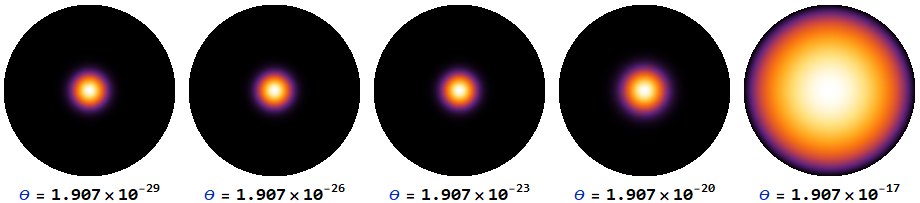}
    \caption{\sf (color online) Effect of $\theta$ on $|\Psi_{0,0}|^2$ at $0 < B\theta < \hbar$ bounded by the radius $r = 2 \times 10^{-9}$~m. 
    %	and $B = 12q$.
    }
    \label{fig:9}
\end{figure}

\subsection{Discussion}

As mentioned in the very first section, $n_r$ is called the radial quantum number and it is responsible to count the number of nodes and antinodes of the radial part of the wavefunctions. For the probability density functions of the ground and excited states, we can only observe the nodes designated as concentric rings due to the rotational symmetry of the eigenfunctions. Hence, when $n_r$ increases so does the number of nodes. It is also apparent that as $|m_l|$ increases, the radii of concentric rings also increase. This implies that this is true regardless of the sign of $m_l$.

On the other hand, when $\theta$ is varied, we only prepare the density plots of the probability distribution functions at the ground state $|\Psi_{0,0}|^2$ since the effect can naturally be extended to higher-order states. At $|\Psi_{0,0}|^2$, the maximum point of these functions are concentrated at their center i.e, the origin designated by the brightest spot on the plots which imply that it is most likely to find the particle there. These Gaussian-like functions are also radially symmetric in their distributions and as $\theta$ increases, the probability distributions spread out farther radially outward which suggest that the likelihood of finding the particle further from the origin increases. This behaviour also persists for higher-order states. When it comes to magnetic field, it simply has the opposite effects on the function in comparison to noncommutativity. We want to note however that the effect of noncommutativity (or magnetic field) on the probability densities are generally non-linear. That is why sometimes, observing gradual or abrupt change as in Figure \ref{fig:9} is quite common for particular choice of the parameters. We do not provide the effect of $B$ at constant $\theta$ in Case III since, for $\theta = 1.907 \times 10^{-29}$m, the effect is so minuscule to be noticed for practical values of $B$ and suffice it to say that it is negligible.

\section{Conclusions}

In this work, we conclude that the energy eigenvalues and degeneracies of the charged harmonic oscillator in 2D noncommmutative space using 2-parameter family of unitarily equivalent irreducible representations of the nilpotent Lie group $G_{NC}$ are unique features of the system as they depend on the subjected magnetic field and noncommutativity. The mathematical structure of the rotationally symmetric energy eigenstates are not affected by these two parameters as they are only factors of the variables. One of these variables of the oscillator i.e, mass of the particle of interest has to be real and hence, we need to impose the following condition i.e, $0 \leq B\theta \leq \hbar$. At $B\theta = 0$ which implies that $B = 0$, the solution of the eigenvalue problem is simply the noncommutative planar harmonic oscillator based on minimal coupling prescription. Degenerate energy levels can be determined when $\theta_d = \bigg\{\left( \frac{k}{\sqrt{n(n+k)}}\right)\frac{\hbar}{m\omega}\bigg\}$ where $n,k$ are positive integers. When it comes to $B\theta = \hbar$, we will end up with an energy spectrum which is isomorphic to Landau problem in symmetric gauge which is very well understood and we know immediately that every energy level is infinitely degenerate regardless of any given values of $B$ and $\theta$ respectively. The remaining case however requires that $\theta_d = \bigg\{\left(f_{exp} - \frac{4nk}{n^2 - k^2}\right)\frac{\hbar}{m\omega}\bigg\}$ or $\theta_d = \bigg\{\left(f_{exp} - \frac{n^2 - k^2}{nk}\right)\frac{\hbar}{m\omega}\bigg\}$ where $n,k$ are coprime and not simultaneously odd. The quantity $f_{exp}$ is a controlling parameter based on \textit{experimental} setup.

For future prospect, further study can be implemented to construct more generalized gauge invariant transformation so that the study of gauge invariant degeneracies and symmetric wavefunctions can be extended to different variants of the problem for example, other exactly solvable eigenvalue problems with different potentials, parameter-dependent eigenvalue problems (e.g, energy-dependent harmonic oscillator, time dependence in mass and frequency, etc.), relativistic models, time evolution, etc.

\section*{Acknowledgments}

NMS wish to acknowledge the Institute for Mathematical Research (INSPEM), Universiti Putra Malaysia (UPM) through Interim Researcher Funding Initiative (INSPEM/IRFI/2/ 2020/6233205) for part of the work.

\end{document}